\documentclass[12pt]{article}

\usepackage{graphicx}

\begin{document}

\begin{center}
{\LARGE Full Bayesian analysis for a class of}

\vspace{.3cm}

{\LARGE jump-diffusion models}
\end{center}

\vspace{.6cm}

 {\bf Laura L.R. Rifo}\\ {\small {\it Universidade Estadual de Campinas, lramos@ime.unicamp.br}}

 {\bf Soledad Torres}\\ {\small {\it Universidad de Valpara\'{\i}so, Soledad.Torres@uv.cl}}

\vspace{.6cm}

\begin{abstract} A new Bayesian significance test is adjusted for jump detection in a diffusion process. This is an advantageous procedure for temporal data having extreme valued outliers, like financial data, pluvial or tectonic forces records and others.

\vspace{.6cm}

\end{abstract}

\section{Introduction}\label{intro}

In the context of financial data some variables often present strong discontinuities, the so-called jumps. The well-known stochastic differential equation driven by a Brownian motion of Black and Scholes \cite{BS} may then becomes unsuitable. 

Decision-making in the presence of jumps has been recently considered, in both theoretical and empirical work.   Merton \cite{Me2} has proposed diffusion models with jumps where the logarithm of jump sizes is assumed to be Gaussian. Kou \cite{K} has suggested a double exponential law for that variable and a more general case, with the power exponential distribution, has been considered by Galea, Ma and Torres \cite{GMT}.

Several inferential techniques have been developed in this area. Lee and Mykland \cite{leemyk} present a nonparametric approach, Chan \cite{chan} suggests maximum likelihood estimation. Continuous time models face however a difficulty in detecting jumps, as available data are obviously discrete. 

On the other hand, diffusion processes with jumps are inherently non-identifiable models: trajectories are sums of diffusion and jump processes. The usual methodology of classical statistical inference then become inappropriate (Khalof, Saphores and Bilodeau \cite{khsabi}; Luan and Xie \cite{luaxie}). Finally, asymptotic sampling distributions are particularly inadequate for small samples in this context.

This paper proposes a full Bayesian inference approach for the problem. The Full Bayesian Significance Test (FBST) was developed by Pereira and Stern \cite{perste} for sharp hypothesis testing in parametric models. The FBST is also the optimal solution for the considered decision problem, as shown by Madruga, Esteves and Wechsler \cite{maeswe}, who obtained well-defined loss functions that make FBST a genuine Bayes test. Such loss functions are very useful in our context as the statistician may fix them to particular numerical descriptions of her world. A deep analysis and revision of FBST may be found in Pereira, Stern and Wechsler \cite{pestwe}.

This paper considers diffusion models with jumps driven by the following stochastic differential equation 
$$
{dS_t \over S_t} = \mu dt + \sigma dW_t + dJ_t =\mu dt + \sigma
dW_t + d\left( \sum_{i=1}^{N_t} (V_i -1)\right).
$$
In the equation above, $W_t$ is a standard Brownian motion, $N_t$ is a Poisson Process, and $V_i$ are non-negative independent identically distributed random variables. The derivative $\mu$ represents expected return, and $\sigma$ represents the volatility.

By making time discrete with unit steps, the equation above can be approximated by the difference equation 
$$
{\triangle S_t \over S_t} = \mu + \sigma Z + B \cdot X .
$$
$B$, $Z$, and $X$ are independent random variables, $B$ and $Z$ having a Bernoulli and a standard Normal law, respectively.

A natural parameterization of the previous convolution allows us to frame the problem of jump detection as a test of hypothesis. In this test, the null hypothesis of no jumps is a sharp hypothesis.

Section 2 describes the formulation of the diffusion model with jumps and its discrete version. Section 3 presents the Full Bayesian Significance Test. The parameterization of the model and the application of FBST to it are seen in Section 4. Section 5 has numerical results for both real and simulated data, yielding also parameter estimates of maximum posterior density. Section 6 discusses possible generalizations and presents conclusions.

\section{The jump-diffusion model formulation }\label{s:jdmf}

In this section we state the jump-diffusion model that motivated
the statistics test that we are interested in. Let
$(\Omega,\mathcal{F},P,\{\mathcal{F}_t\})$ be a completed filtered
probability space on which is defined a Brownian motion $W$ and a
compound Poisson process $J$, both adapted to the filtration
$\{\mathcal{F}_t\}$. More precisely, we assume that the process
$J$ takes the following form:
\begin{equation}
J_t=\sum_{j=1}^{N_t}(V_j-1), \quad t\ge 0,
\end{equation}
where $N= \{N_t\}$ is a standard Poisson process with rate
$\lambda$, and $\{V_j\}$is a sequence of i.i.d. nonnegative random
variables. We assume that:
\begin{enumerate}
    \item for each $j$, $X_j=log(V_j)$ has a given distribution;
    \item the process $W$, $N$, and $X_j$'s are independent;
    \item $\mathcal{F}_t=\sigma\{W_s,J_s: 0 \le s \le t\}, t\ge
    0$, augmented under $P$ so that it satisfies the \it{usual
    hypothesis}.
\end{enumerate}
In our jump-diffusion model we assume that all economics have a
finite horizon $[0,T]$, and the price of our underlying risky
asset is given by the following stochastic differential equation:

\begin{equation}\label{Precio}
{ dS_t \over S_t }= \mu dt + \sigma dW_t + dJ_t =\mu dt + \sigma
dW_t + d\left( \sum_{i=1}^{N_t} (V_i -1)\right).
\end{equation}

For notational simplicity and in order to get analytical solutions, the drift $\mu$ and the volatility $\sigma$ are assumed to be constants, and the Brownian motion and jumps are assumed to be one dimensional. These assumptions, however, can be omitted to develop a general theory.

\subsection{Discrete Model}

The goal of this section is to approximate the equation given in
(\ref{Precio}) using the Euler method. We know that, from Protter \cite{P},
the solution to the Stochastic Differential Equation
(\ref{Precio}), that give us the dynamics of the asset price, is given by
\begin{equation}\label{solucionprecio}
S_t = S_0 e^{\left(\mu - {\sigma ^2 \over 2} \right)t + \sigma
W_t} \prod_{i=1}^{N_t} V_i.
\end{equation}
Next,
\begin{eqnarray}\label{deltaprecio}
{ \triangle S_t \over S_t} & = & {S_{t+1} - S_t \over S_t }\nonumber\\
& = & exp
\left\{ \left(\mu - {\sigma ^2 \over 2} \right)\triangle t +
\sigma \left( W_{t+ \triangle t} - W_t \right) +\displaystyle
\sum_{i=N_t + 1}^{N_{t + \triangle t}} X_i \right\} -1 .
\end{eqnarray}
If $\triangle t$ is small enough, we can reject the terms of
greatest order from the Taylor expansion, approximating $e^x$ by
$1 +x+ x^2/2$. We obtain
$$
{ \triangle S_t \over S_t} \sim  \left(\mu - {\sigma ^2 \over 2}
\right)\triangle t+ \sigma \left( W_{t+ \triangle t} - W_t \right)
+\displaystyle \sum_{i=N_t + 1}^{N_{t + \triangle t}} X_i + {1
\over 2} \sigma^2 \left( W_{t+ \triangle t} - W_t \right)^2
$$
\begin{eqnarray}\label{deltaaproxprecio}
&\sim & \mu \triangle t + \sigma Z \sqrt{\triangle t} +
\displaystyle \sum_{i=N_t + 1}^{N_{t + \triangle t}} X_i,
\end{eqnarray}
where $Z$ is a normal standard random variable and the unknown parameters
 are $\mu$, which represents the expected return,
$\sigma$, the volatility, and $\lambda$, the jump rate.

As it was shown in Kou \cite{K}, for $\triangle _i$ small enough we
have:
\begin{equation}\label{precioapp}
\sum_{i=N_t + 1}^{N_{t + \triangle t}} X_i =
\left\{%
\begin{array}{ll}
    X_{N_t+\triangle}, & \hbox{w.p. $\lambda \triangle$;} \\
    0, & \hbox{w.p. $1-\lambda \triangle$.} \\
\end{array}%
\right.
\end{equation}
In other words, if $\delta = |\pi|$ is sufficiently small, the
return can be approximated in distribution by
\begin{equation}\label{returnapp}
{\triangle S_t \over S_t} = \mu \delta + \sigma Z \sqrt{\delta} +
B \cdot X ,
\end{equation}
where $B$ is a Bernoulli random variable with $P(B=1)= \lambda
\delta$ and $P(B=0)= 1- \lambda \delta$, and $Z \sim N(0,1)$. Note
that
$$
\hspace*{-6cm} P(\sigma \sqrt{\delta} Z + BX \le x) =
$$

\vspace{-1cm}

\begin{eqnarray}\label{SumaPP}
 &=& P(\sigma \sqrt{\delta} Z
+ X \le x)  P(B=1) +  P(\sigma \sqrt{\delta} Z\le x) P(B=0)
\nonumber \\
&=& P(\sigma \sqrt{\delta} Z + X \le x)\lambda \delta + P(\sigma
\sqrt{\delta} Z\le x) (1-\lambda \delta).
\end{eqnarray}
So, returns can be modeled as the convolution $\sigma \sqrt{\delta} Z + B\,X$, between independent normal random variable and a random variable with some distribution $F$.

In this work, we will assume that $\delta =1$ and random variable $X$ has Bernoulli distribution with unknown parameter $p$. Then, $BX$ has a Bernoulli distribution given by
\begin{equation}\label{bernoulli}
f(x)=\left\{%
\begin{array}{ll}
    \lambda p & \hbox{for $x=1$;} \\
    1-\lambda p, & \hbox{for $x=0$.} \\
\end{array}%
\right.
\end{equation}
There are many alternative models depending on the distribution of the jumps, see for example Merton \cite{Me2}, for the normal distribution, Kou for double exponential, and Galea, Ma and Torres \cite{GMT} for a generalization of the previous works.

From (\ref{SumaPP}) and (\ref{bernoulli}), the density function of $\mu + \sigma Z + BX$ has the following representation:
    \begin{equation}\label{NG}
    f(y) = {1 \over \sqrt{2\pi}\sigma }\left((1-\lambda p) e^{{-1 \over 2\sigma^2}(y-\mu)^2}
    + \lambda p\,e^{{-1 \over 2\sigma^2}(y-(\mu+1))^2}\right) .
    \end{equation}

\begin{figure}[!htp]
\hfill
\includegraphics[scale=.3]{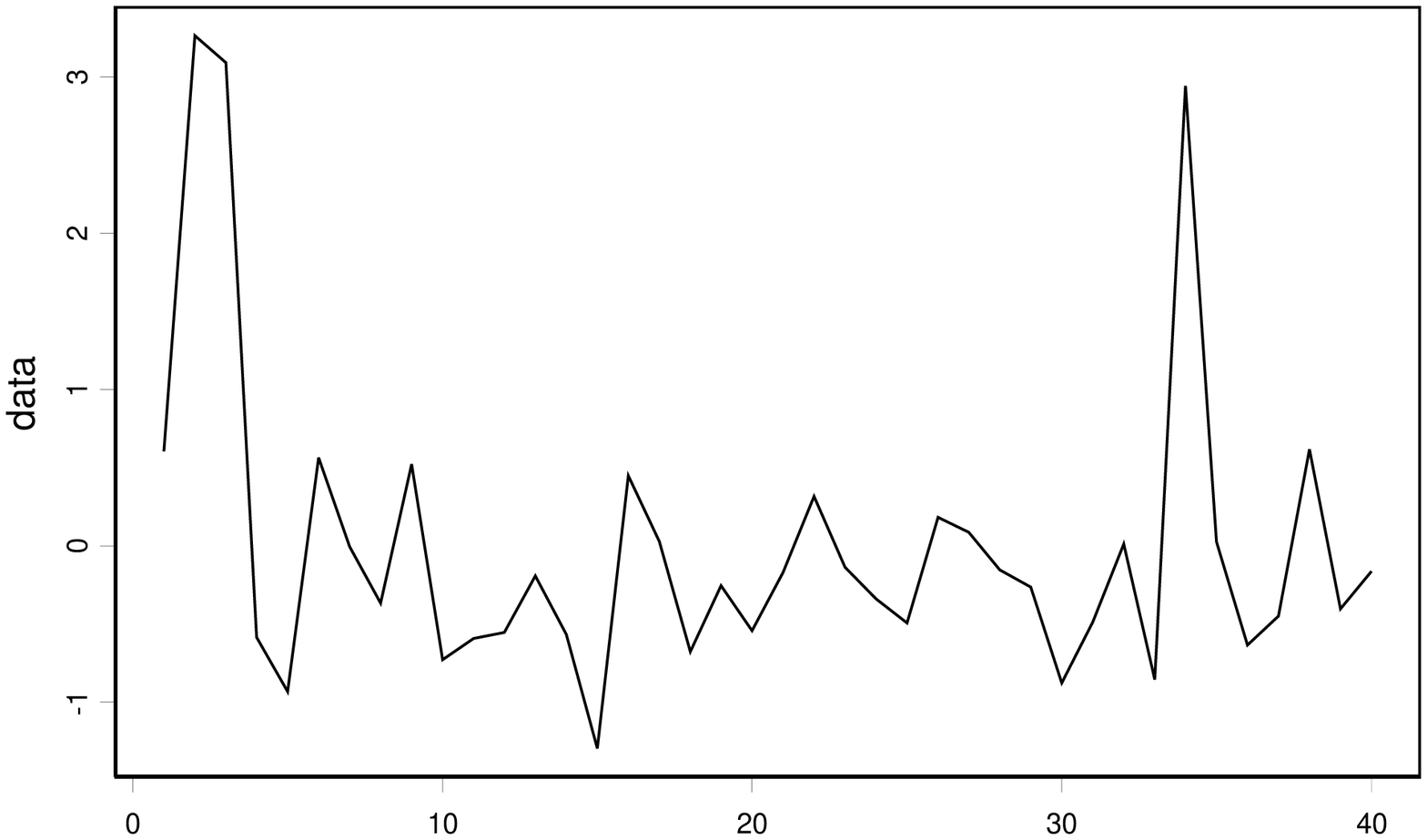}
\includegraphics[scale=.3]{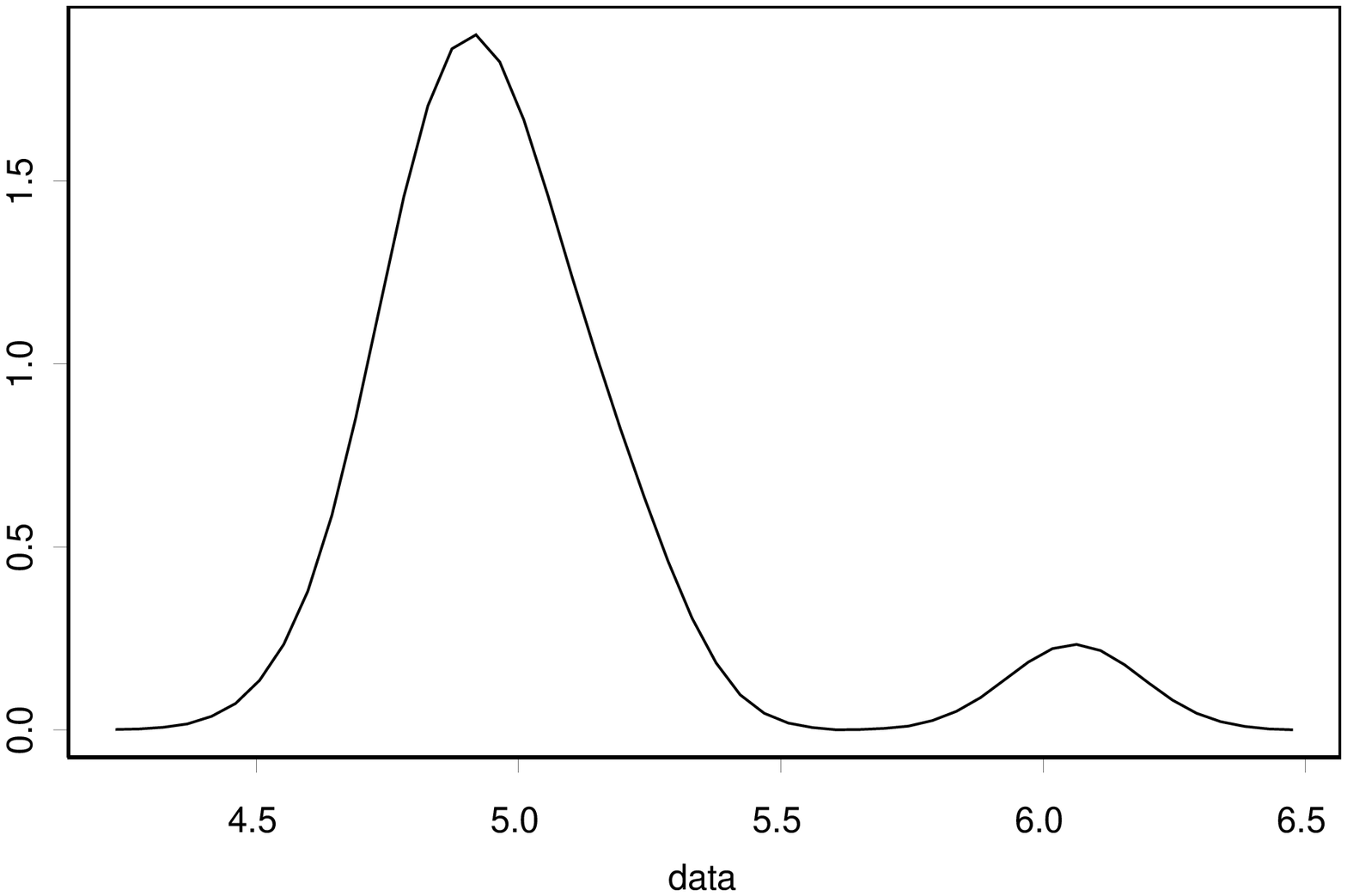}
\hfill
\vspace{-.5cm}
\caption{{\small Time plot and smoothed density of simulated data, $\lambda p=0.05$.}} \label{figure1}
\end{figure}

Figure \ref{figure1} shows a realization of the returns $\Delta S_t/S_t$ and the density (\ref{NG}), for $\Delta t=1$ day, $\mu = 0$, $\sigma = 0.2$, $\lambda p=0.05$. In other words, for this process, there are about $\lambda p = 5\%$ jumps per year with size one.

\section{Full Bayesian Significance Test}\label{s:fbst}

Let us consider a random variable D whose value $d$ in the measurable sample space $(\Omega, \mathcal{S})$ is to be observed.

Let $\Theta$ be the parametric space, that is, a set such that $Pr(\cdot | \theta)$ is a well-defined probability measure in $\mathcal{S}$, for all $\theta \in \Theta$. Denote by $(\Theta, \mathcal{B} , \pi)$ a probability measure structure on $\Theta$ such that $\pi$ determines a priori probability distribution on $\Theta$.

After observing data $d$, the information about $\theta$ is updated by Bayes' theorem and quantified by the posterior probability law on $\Theta$, $\pi_d$.

Full Bayesian Significance Test (FBST) procedure is defined in the case when this posterior distribution has a density function with respect to Lebesgue measure.

Let $f(\theta)$ denote the priori density, $f(d | \theta)$, the likelihood function of $\theta$ after observing data $d$, and $f(\theta| d)$, the posterior density of $\theta$ given data $d$, related by
$$
f(\theta| d) \propto f(\theta) f(d | \theta) .
$$

A precise hypothesis $H$ can be defined as a submanifold $\Theta_0 \subset \Theta$ such that $dim (\Theta_0) < dim (\Theta)$. This implies that the posterior probability of a precise hypothesis is null for an absolutely continuous posterior distribution: every precise hypothesis should be rejected in that case.

In order to avoid such a drastic conclusion, FBST deals not directly with $\Theta_0$, but with a sort of critical region defined by the level surfaces of the posterior density $f(\theta| d)$.

Let us define the {\it tangential set} $T_0$ to the null hypothesis $\Theta_0$ as the set
\begin{equation}\label{eq:t0}
T_0 = \{ \theta \in \Theta : f(\theta | d)> f_0 \} , 
\qquad \mbox{ where } f_0 = \sup_{\Theta_0} f(\theta | d) .
\end{equation}
In other words, the tangential set to $\Theta_0$ considers all points ``most probable'' than $\Theta_0$, according to the posterior law.

The posterior probability of $T_0$, 
\begin{equation}\label{eq:k0}
\kappa_0 = \int_{T_0} f(\theta | d) d\theta ,
\end{equation}
is called its {\em credibility}. The {\it evidence} for the null hypothesis is then defined as
\begin{equation}\label{eq:ev}
ev (\Theta_0 ) = 1 - \kappa_0 = 1 - \pi_d (T_0) .
\end{equation}

So, if tangential set has high posterior probability, the evidence in favor of $\Theta_0$ is small; if it has low posterior probability, the evidence against $\Theta_0$ is small.

In Madruga et al. \cite{maeswe}, the Bayesianity of the test of significance based on this evidence measure is showed, in the sense that there exists a loss function such that the decision for rejecting the null hypothesis is based on its posterior expected value minimization.

The computation of $ev (\Theta_0 )$ can be performed in two steps: a numerical optimization procedure to find $f_0$, and a numerical integration to find $\kappa_0$.

\section{FBST for Bernoulli jumps}\label{s:fbstbj}

For the problem presented in Section \ref{s:jdmf}, let us consider the parametric space
$$
\Theta = \{ \theta = (\lambda, p, \mu, \sigma^2)\in R^4 : \lambda, p \in [0,1], \mu \in R, \sigma^2>0\} .
$$ 

According to equation (\ref{NG}), the likelihood for the parameter $\theta$, given a sample $d=(x_1,\dots, x_n)$, may be written as
\begin{eqnarray}\label{eq:likelihood}
f(d |\theta) & = & \prod_{i=1}^n \frac{1}{\sqrt{2\pi\sigma^2}}  \left[ \lambda p\, e^{-\frac{1}{2\sigma^2}(x_i-(\mu+1))^2}+(1-\lambda p)e^{-\frac{1}{2\sigma^2}(x_i-\mu)^2}  \right] \nonumber\\
& = & (2\pi\sigma^2)^{-n/2} e^{-\frac{1}{2\sigma^2}\sum_{i=1}^n (x_{i}-\mu)^2 } \\
& & \times \sum_{k=0}^n (\lambda p\, e^{-\frac{1+2\mu}{2\sigma^2}})^k (1-\lambda p)^{n-k}\sum e^{\frac{1}{\sigma^2}\sum_{l=1}^k x_{i_l}} .\nonumber
\end{eqnarray}

From the above equation, we note that if $\lambda_1\, p_1=\lambda_2\, p_2$, then the parameters $(\lambda_1, p_1, \mu, \sigma^2)$ and $(\lambda_2, p_2, \mu, \sigma^2)$ are observationally equivalent, as they have the same likelihood from data $d$. 

The function $I:\Theta \to \cal{T}$, $I(\lambda, p, \mu, \sigma^2)=(\lambda p, \mu, \sigma^2)$ is an identifying function, that is, it is a one-to-one transformation under that equivalence relation. (See Kadane \cite{kada} for more properties and examples of identifying functions.)

Theorem 5 in Kadane \cite{kada} states that the Bayesian analysis can be done on $\cal{T}$, from the prior induced on $\cal{T}$ by $I$ and the likelihood function.

Aiming to preserve the information of an experiment about the parameters of interest, we will define a prior distribution for $\theta$ such that $f(\lambda, p, \mu, \sigma^2)=f(\lambda', p', \mu, \sigma^2)$ whenever $\lambda \, p = \lambda' \, p'$.

It will be assumed that $(\lambda,p)$ and $(\mu, \sigma^2)$ are independent random vectors, yielding
$$
f(\lambda, p, \mu, \sigma^2) = f(\lambda, p)f(\mu, \sigma^2),
$$
for every $(\lambda, p, \mu, \sigma^2)\in\Theta$.

The following prior density will be adopted for $(\lambda,p)$
$$
f(\lambda, p \,| \beta) \propto (1-\lambda p)^{\beta-1} \qquad \mbox{para} \quad 0\leq \lambda, p\leq 1 ,
$$
with known $\beta >0$.

In order to ease some numerical aspects, we will assume a uniform on $(0,\sigma^2_0)$ density for $\sigma^2$ and, given $\sigma^2$, a uniform on $(\mu_0-c\sigma,\mu_0+c\sigma)$ law for $\mu$, with hyperparameters $\sigma^2_0 , c >0$, $\mu_0 \in R$.

The posterior density for $\theta = (\lambda, p, \mu,\sigma^2)$ is then
\begin{eqnarray}\label{eq:posterior}
f(\theta | d) & \propto & f(d|\theta) f(\lambda , p)f(\mu, \sigma^2) \nonumber \\
& \propto &  f(d|\theta) (1-\lambda p)^{\beta -1}\frac{1}{\sigma}\; 1_{(\mu_0 -c\sigma , \,\mu_0 + c\sigma)\times (0,\sigma_0^2)} (\mu ,\sigma^2) .
\end{eqnarray}

The null hypothesis,
\begin{equation}\label{eq:H0}
\Theta_0 = \{ (\lambda ,p, \mu,\sigma^2)\in \Theta : \lambda  p=0 \} ,
\end{equation}
states that the process has no jumps, a sharp hypothesis.

Under this formulation, the measure of evidence in favor of $\Theta_0$, $ev(\Theta_0)$, defined by (\ref{eq:ev}), allows us to perform a significance test for $\Theta_0 $ without having to modify the posterior distribution (either by assigning positive probability to the null hypothesis or by enlarging it, both polemical solutions in the literature).

\section{Numerical results} \label{s:nr}

\subsection{Simulation}

We used S-Plus to perform FBST on simulated data from the convolution of a Normal$(5,0.2^2)$ and a Bernoulli with parameter $\lambda p$ having the values 0, 0.025, 0.10, 0.35 and 0.50.

Data were standardized
$$
z_i = \frac{x_i-\bar{x}}{s_x} \quad , i=1,\dots , n ,
$$
where $\bar{x}$ and $s_x$ are, respectively, the sample mean and sample standard deviation, with sample size $n=40$. The hyperparameters for the prior distribution were chosen to be $\beta =1$, $\sigma_0 = 10$, $\mu_0=0$, $c=10$.

\begin{table}[ht]
\begin{small}
\begin{center}
\begin{tabular}{cccc}
$\lambda p$ & $ev(\Theta_0)$ & posterior mode $(\lambda p,\mu,\sigma)$ & posterior mean $(\lambda p,\mu,\sigma)$ \cr
\hline
0 & 1 & (.0000, 5.045, .2087) & (.4077, 4.629, .2139)  \cr
0.025 & .00382 & (.06211, 4.993, .2071) & (.08346, 5.005, .2044) \cr
0.10 & 3.98e-8 & (.1494, 5.002, .1510) & (.1438, 5.011, 0.1766) \cr
0.35 & 3.47e-5 & (.3416,5.029,.1719) & (.3501, 5.021, .1933) \cr
0.50 & 3.37e-6 & (.5193, 5.046, .1826) & (.5184, 5.049, .2046) \cr
\end{tabular}
\end{center}
\end{small}
\vspace{-.4cm}
\caption{{\small In the second column, evidence values for several jump rates, $\lambda p$, and modal and mean estimates for $\theta'$ at the third and fourth column, respectively.}}\label{t:1}
\end{table}

Table \ref{t:1} presents the evidence values $ev(\Theta_0)$ for the distinct rates simulated, and also maximum posterior estimates and the mean posterior estimate for $\theta'=(\lambda p,\mu,\sigma)$.

\begin{figure}[ht]
\begin{center}
\includegraphics[scale=.3]{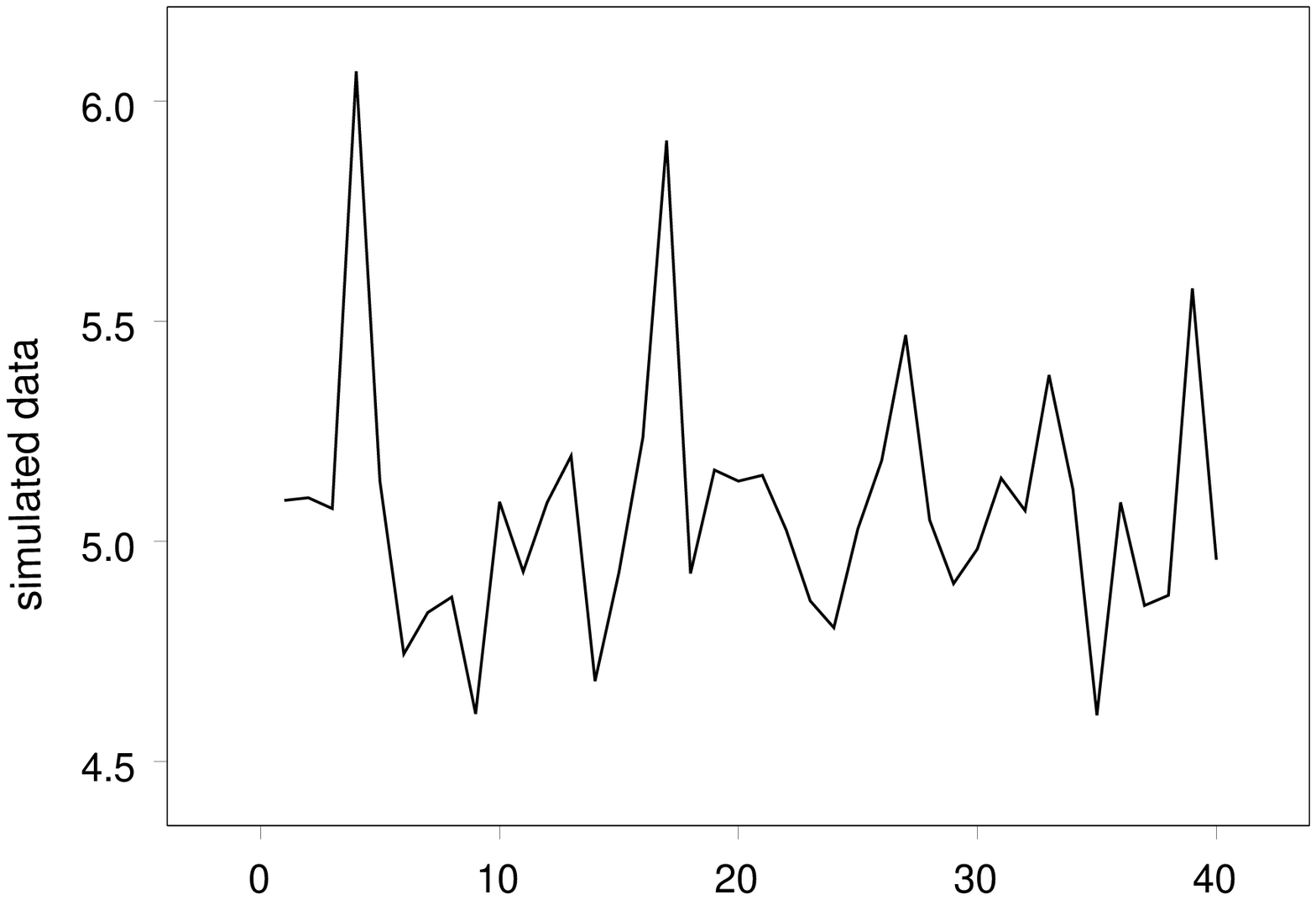}
\includegraphics[scale=.3]{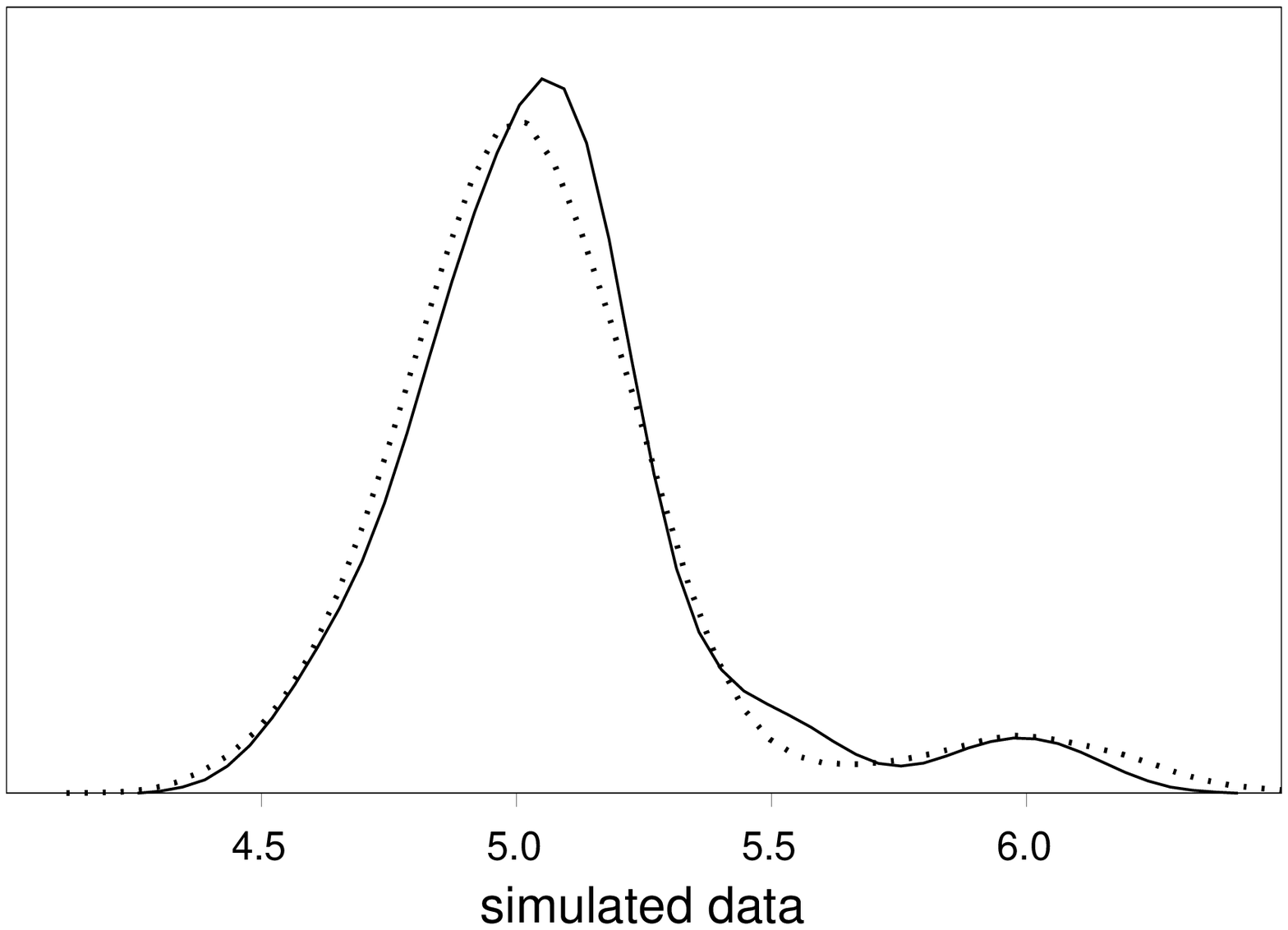}
\end{center}
\vspace{-.7cm}
\caption{{\small Time plot and fitted density for frequencies of simulated data, $\lambda p = 0.025$, $ev (\Theta_0 )=.003817$.}} \label{figure2}
\end{figure}

Figure \ref{figure2} presents the simulated data on time for a process having $\lambda p = 0.025$ as jump rate. The right-hand graph shows the data empirical density (in full line), and the density adjusted by the posterior mode (in dotted line). Data suggest the possibility of jumps, quantified by the small evidence in favor of the null hypothesis, $ev (\Theta_0 )=.00382$.

\begin{figure}[ht]
\begin{center}
\includegraphics[scale=.3]{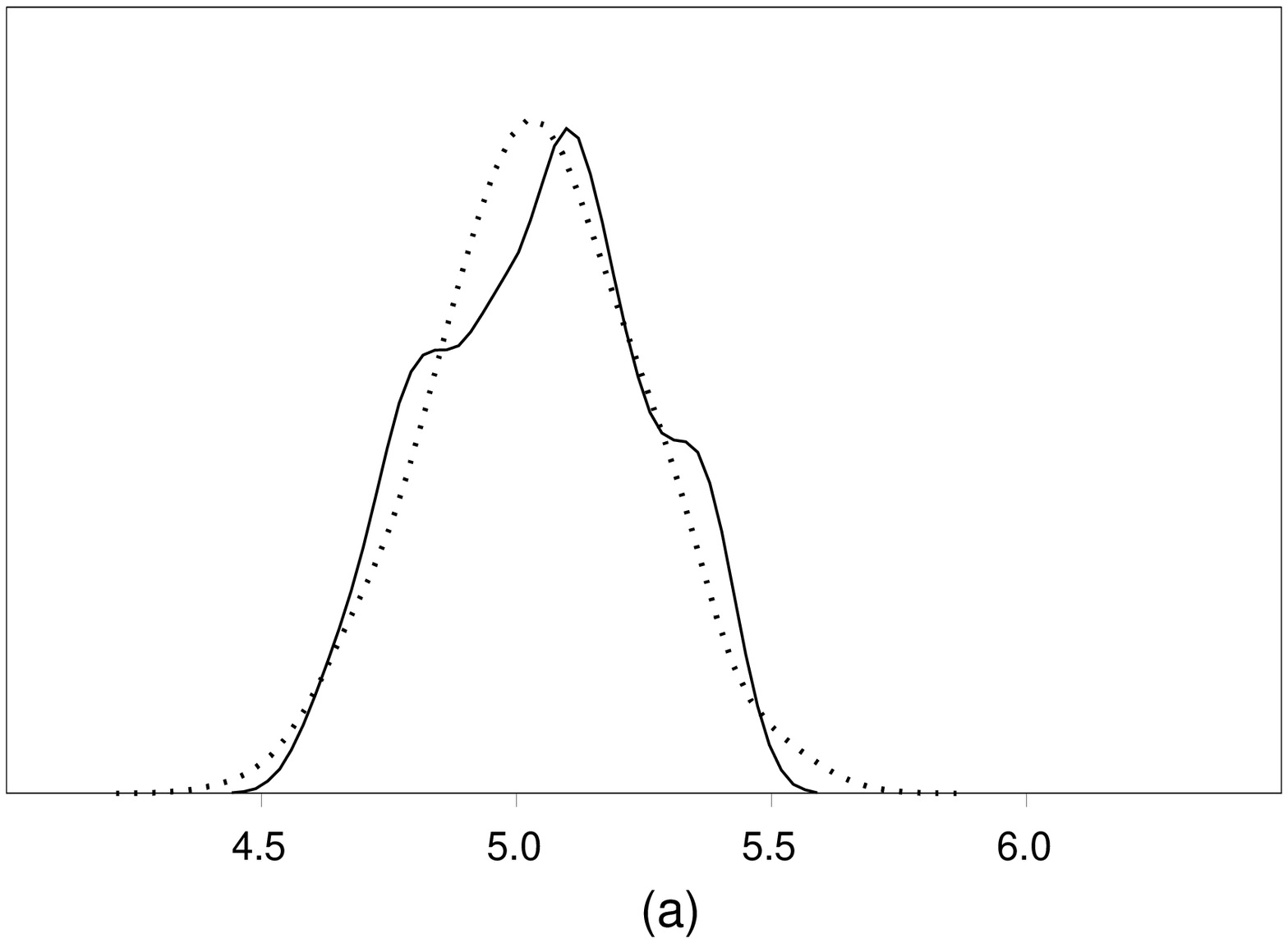}
\includegraphics[scale=.3]{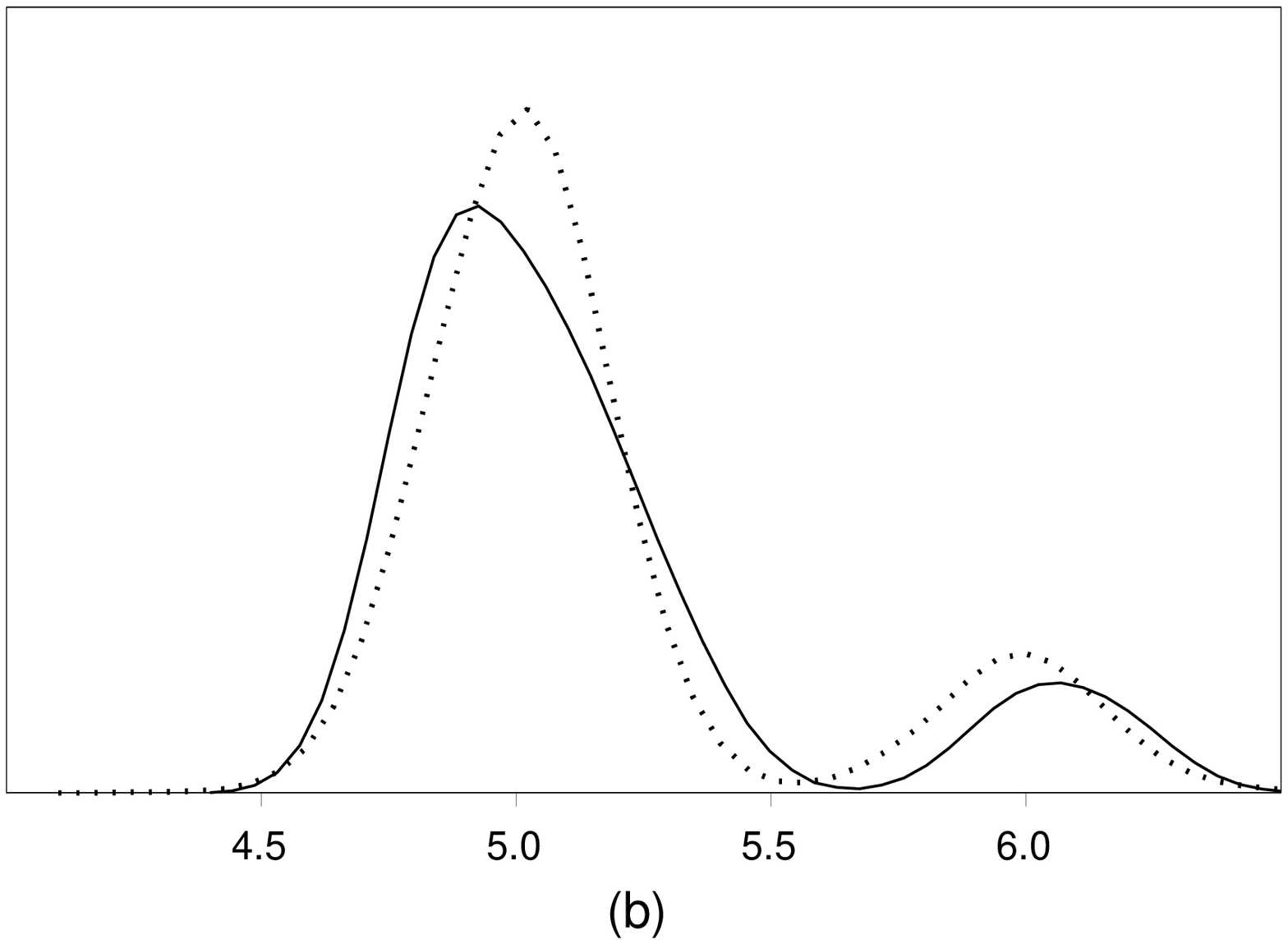}
\includegraphics[scale=.3]{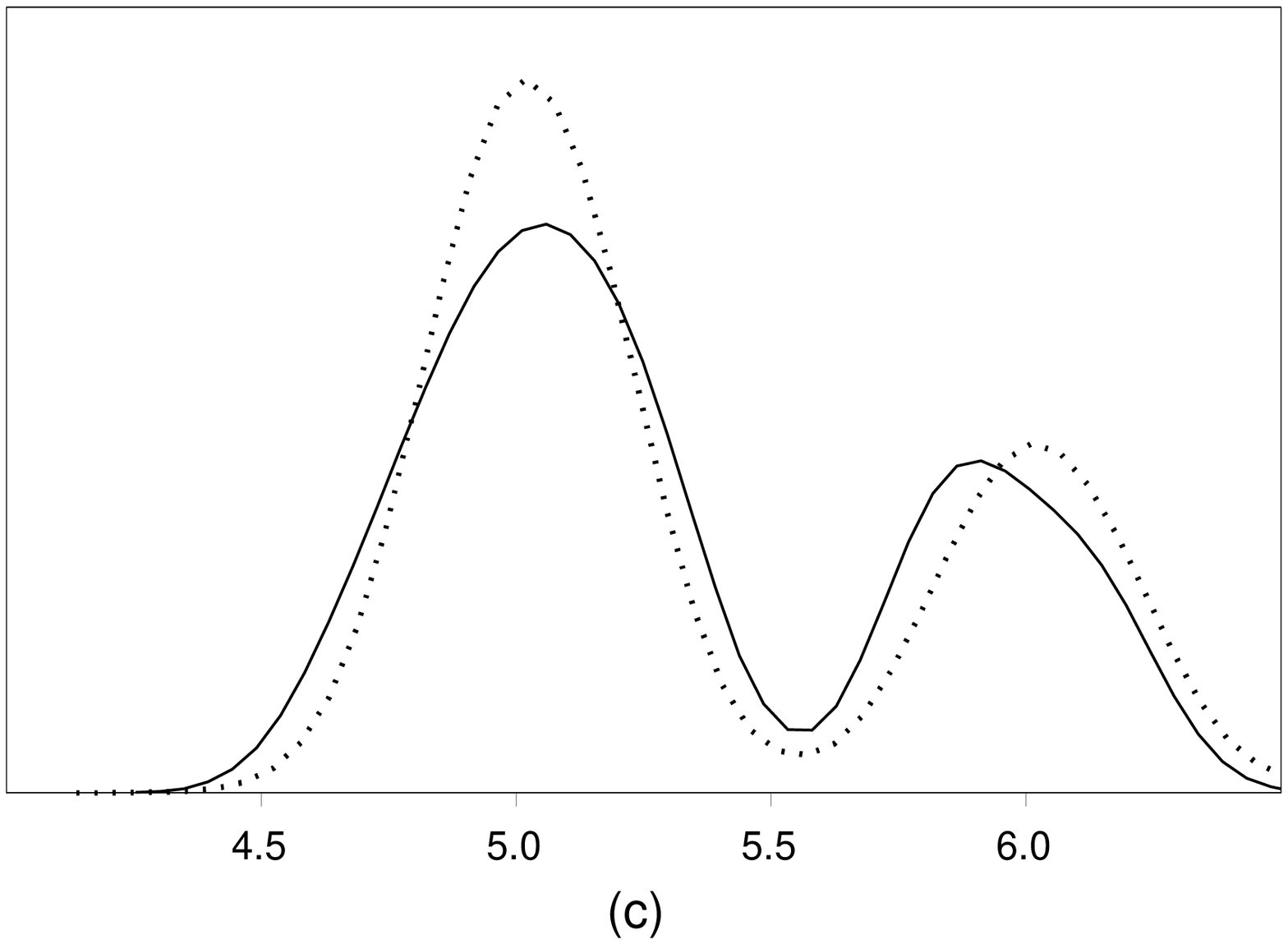}
\includegraphics[scale=.3]{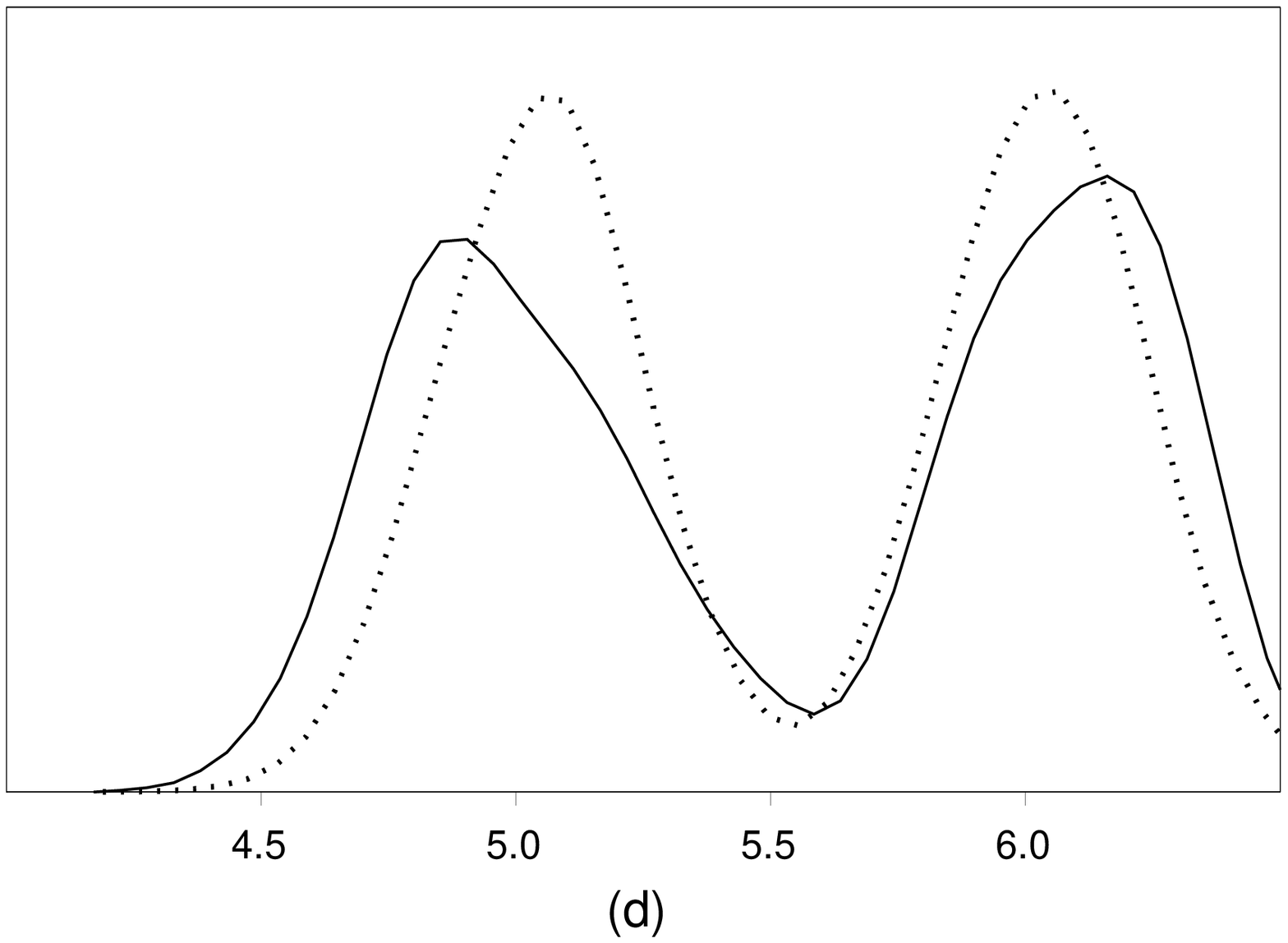}
\end{center}
\vspace{-.8cm}
\caption{{\small Simulated Data empirical density (full line), density adjusted by posterior mode (dotted line), with jump rate and evidence, respectively: (a) $\lambda p=0$, $ev (\Theta_0 )=1$; (b) $\lambda p=0.1$, $ev (\Theta_0 )=.004662$; (c) $\lambda p=0.35$, $ev (\Theta_0 )=3.47e-5$; (d) $\lambda p=0.5$, $ev (\Theta_0 )=3.37e-6$.}}
\label{figure3}
\end{figure}

Graph \ref{figure3} shows the simulated data empirical density (full line) in each case and the density determined by the posterior mode (dotted line), for the various rates considered.

It can be seen, a strong bimodality in the sample is well detected by the FBST procedure as expected. On the other hand, without a strong bimodality, the evidence in favour of $\Theta_0$ is large, as seen in the graph of simulated data with $\lambda p=0$, where $ev (\Theta_0 )=1$.

The computation of evidence values $ev (\Theta_0 )$ was based on the simulation of 400,000 independent points uniformly distributed over the support of the posterior distribution, and having their non-normalized values, $\pi(\theta|d)$, computed. Other 400,000 points were uniformly generated on $\Theta_0$, in order to obtain the sample maximum of the non-normalized posterior density on $\Theta_0$, $\varphi_0$.

The posterior probability of the tangential set can be approximated by the ratio of the sum of values of $\pi(\theta|d)$ larger than $\varphi_0$ to the total sum of $\pi(\theta|d)$ for all generated points.

The mean time for running of this program was 600 seconds approximately, in a standard PC.

Other optimization numerical methods which are more efficient for determining $\varphi_0$ and integrating on $T_0$, may be used. Our simple method, however, had a good performance for the problem of model choice between models having Bernoulli jumps or models without jumps.

Furthermore, it can be generalized in a straightforward way for other jump.

\subsection{Real data}\label{s:real}

\begin{figure}[!htp]
\hfill{}
\includegraphics[scale=.3]{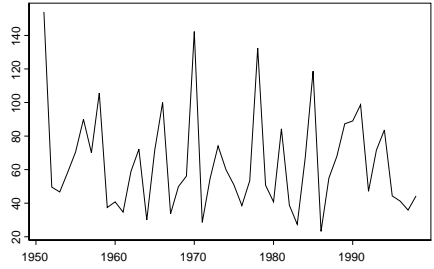}
\hfill{}
\vspace{-.5cm}
\caption{{\small Venezuela Annual Maximum Rainfall Data, 1951 to 1998.}} \label{figure4}
\end{figure}

Figure \ref{figure4} presents Annual Maximum Rainfall Data (Maiquetia station at Venezuela central coast) during 1951-1998. Source:
\begin{quote}
http://www.blackwellpublishing.com/rss/Volumes/Cv52p4.htm
\end{quote}

Coles and Pericchi \cite{colper} use these data to compare Bayesian and Classical methods on the prediction of the 1999 catastrophic maximum of 410mm.

Data were transformed to the standard scale, as described in the previous section. The jump size was considered random, so the parametric space had one more dimension added to it, $K$, having a uniform prior distribution on $[0,30]$, independently from the other components.

The analysis obtained an evidence value in favour of $\Theta_0$, $ev (\Theta_0 )=.03519$. The posterior mode and mean estimates are
$$
(\hat{\lambda p},\hat{\mu},\hat{\sigma},\hat{k})_{mode} = (0.1447, -0.2682, 0.6130, 2.227) ,
$$
$$
(\hat{\lambda p},\hat{\mu},\hat{\sigma},\hat{k})_{mean} = (0.1683, -0.2831, 0.6576, 2.181) .
$$

In original units, the posterior mode estimate represents the model
$$
\hat{R}_{mode} = \bar{R} - 0.27 sd(R) + 0.61 sd(R) Z + 2.2 sd(R) B_1,
$$
and the posterior mean estimate represents the model
$$
\hat{R}_{mean} = \bar{R} - 0.28 sd(R) + 0.66 sd(R) Z + 2.2 sd(R) B_2,
$$
where $\bar{R}$ and $sd(R)$ are the sample mean and the sample standard deviation, $Z$ has standard normal distribution, $B_1$ has a Bernoulli(.145) distribution and $B_2$, has a Bernoulli(.168) law.

\begin{figure}[!htp]
\hfill{}
\includegraphics[scale=.3]{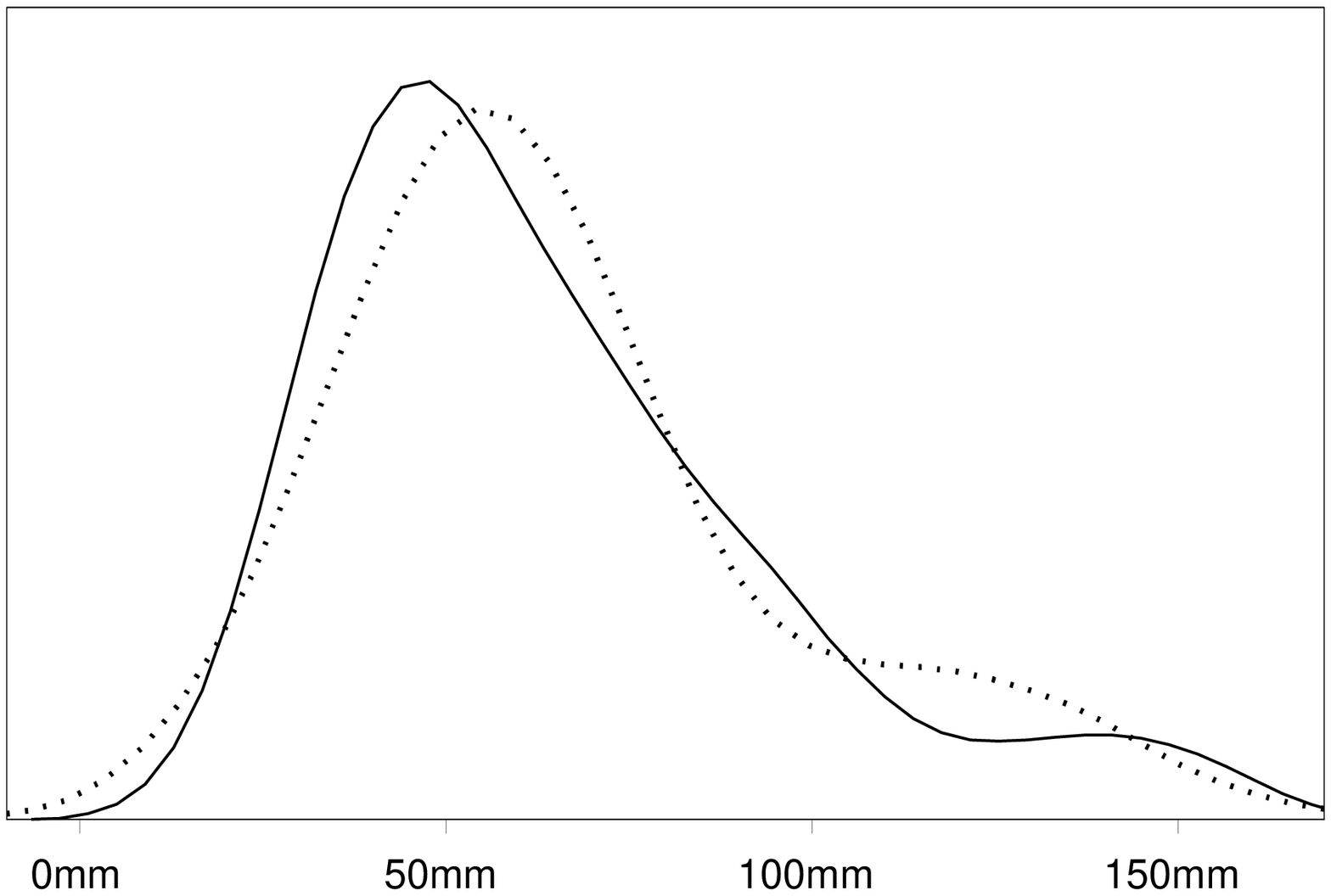}
\includegraphics[scale=.3]{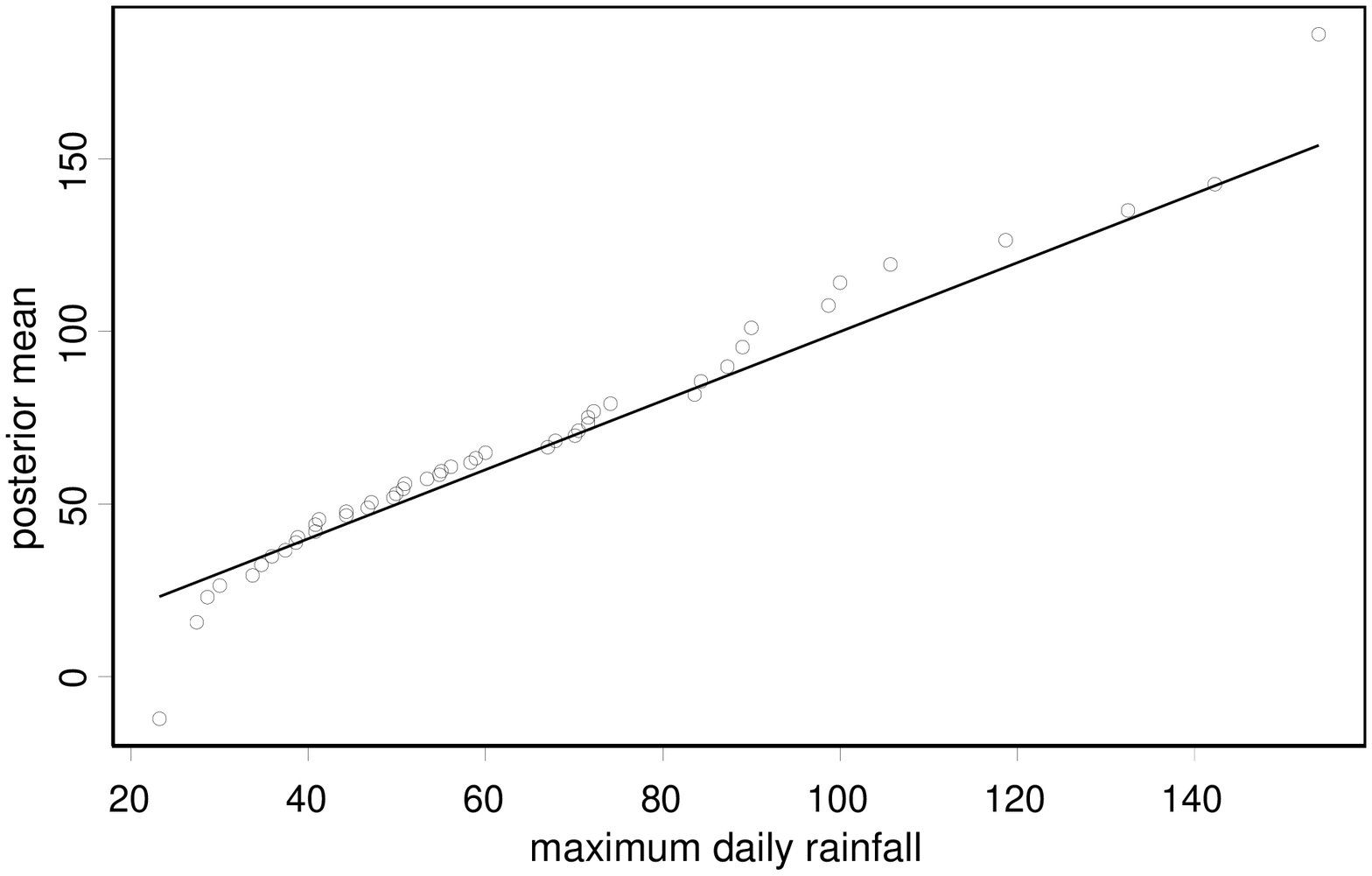}
\hfill{}
\vspace{-.4cm}
\caption{{\small Real Data Empirical Density (full line), posterior mean adjusted density (dotted line); qqplot for data and posterior mean model. Evidence in favour of non-existence of jumps is $ev (\Theta_0 )=0.0351866$.}} \label{figure5}
\end{figure}

As seen in Figure \ref{figure5}, that estimate fits well the observed data.

The posterior mode estimate is here the maximum likelihood estimate, as the chosen prior is uniform.

The posterior distribution simulation yields other relevant probabilities: given a threshold $l$, let $A_t$ be the event that values not smaller than $l$ are recorded only after time $t$. In the example, $A_t$ indicates that an annual maximum rainfall largest than $l$ takes at least $t$ years to be recorded.

$$
P(A_t | d) = \int_{\Theta} P( A_t | \theta) f(\theta | d) d\theta 
$$
can be approximated essentially with the same values obtained by the previous simulation.

\begin{figure}[!htpc]
\hfill{}
\includegraphics[scale=.4]{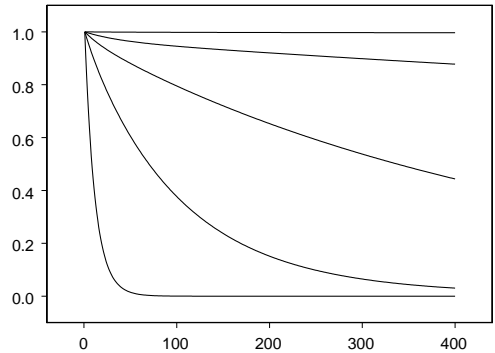}
\hfill{}
\vspace{-.5cm}
\caption{{\small Survival probability until time $t$, for thresholds $q(90)=100$, $q(98)=150$, 165, 180, 350, as a function of time $t$, in years.}} \label{figure6}
\end{figure}

Figure \ref{figure6} shows this probability as a function of time $t$ up to 400 years, for several thresholds. The curves correspond to threshold of 100, 150,  165, 180 and 350mm, respectively from left to right, for annual maximum rainfall.

As an illustration, a maximum value larger than 100mm, that corresponds to the ninth observed decile, has a 10$\%$ chance of being recorded after 27 years, and a 56$\%$ chance of being observed during the next ten years. An annual maximum larger than 165mm, not recorded in the sample, has a 3$\%$ chance of being recorded during the next 10 years, 12$\%$ during the next 50 years and 20$\%$ probability of being observed during the next 100 years.

\begin{figure}[!htpc]
\hfill{}
\includegraphics[scale=.4]{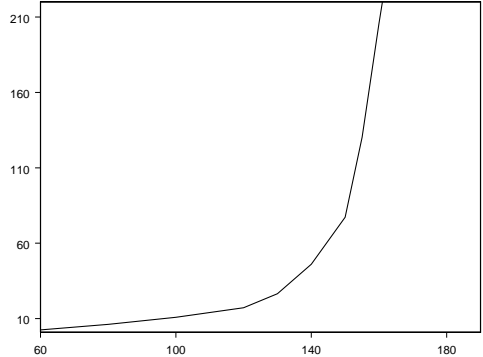}
\hfill{}
\vspace{-.5cm}
\caption{{\small Expected time, $y$ axis, to reach threshold $l$, $x$ axis.}} 
\label{figure7}
\end{figure}

Finally, given a threshold $l$, we may demand the expected time until recording a value not smaller than $l$, $T_l$:
$$
T_l = \frac{1}{P(S \geq l | d)} .
$$

Figure \ref{figure7} presents Expected times $T_l$, to reach threshold $l$ (indicated as abscisse). For instance, the expected time to record an annual maximum higher than 140mm is around 46 years, and 77 years is the expected time until an annual maximum surpasses 150mm.

\section{Conclusions}

In this  work we are testing two kinds of models: diffusion processes versus diffusion processes with Bernoulli jumps. We could also be interested in testing for more general jump families, which allow, for instance, heavy-tailed distributions or diffusion processes having non-constant volatility against processes having constant volatility.

The proposed test procedure extends naturally for each of theses families of models, as long as there is a sharp hypothesis to be tested. Modifications on the likelihood function are straightforward. The computational cost increases with the size of the parameter space, as the required probabilities are integrated directly on the latter.

Noninformative prior distributions may jeopardize mixture properties of the envolved Markov chains and therefore slow down the convergence rate of other Monte Carlo methodologies.

We stress the absolute continuity of the posterior distribution. The chosen priors are usually Lebesgue absolutely continuous, entailing absolute continuity of the posterior for the models in our context.

Any sharp hypothesis has therefore null posterior probability and posterior ratios are not a good criterion to compare models defined by sharp hypotheses.

The measure of evidence used in this paper is associated (Section \ref{intro}) to a decision problem. In the real decision problem, we could - and should - consider the loss function the minimization of its expected value is tantamount to performing the FBST. 

The ideas behind FBST are intuitive and easy to understand. Moreover, non-identifiability does not present a problem for the Bayesian operation.

Finally, the Bayesian approach allows us to answer questions such as those risen in Section \ref{s:real}, with real data, in terms of posterior probabilities and not limited to mere  parametres estimates.

\end{document}